\documentclass[12pt]{JHEP3}

\preprint{  }

\usepackage{epsfig,multicol}
\usepackage{amsmath}
\usepackage{graphics}
\usepackage{graphicx}
\usepackage{dcolumn}
\usepackage{amssymb}
\usepackage{amsthm}
\usepackage{amsfonts}
\usepackage{subfigure}
\usepackage{setspace}

\title{General proof of the entropy principle for self-gravitating fluid in $f(R)$ Gravity}
\author{Xiongjun Fang$^a$, Minyong Guo$^b$ and Jiliang Jing$^a$\footnote{Corresponding author. Email:
jljing@hunnu.edu.cn}
\\ $^a$Department of Physics, and Key Laboratory of Low Dimensional Quantum Structures and Quantum Control of Ministry of Education, Hunan Normal University, Changsha, Hunan 410081, P. R. China\\
$^b$Department of Physics, Beijing Normal University, Beijing 100875, P. R. China}

\abstract{

The discussions on the connection between gravity and thermodynamics attract much attention recently. We consider a static self-gravitating perfect fluid system in $f(R)$ gravity, which is an important theory could explain the accelerated expansion of the universe. We first show that the Tolman-Oppenheimer-Volkoff equation of $f(R)$ theories can be obtained by thermodynamical method in spherical symmetric spacetime. Then we prove that the maximum entropy principle is also valid for $f(R)$ gravity in general static spacetimes beyond spherical symmetry. The result shows that if the constraint equation is satisfied and the temperature of fluid obeys Tolmans law, the extrema of total entropy implies other components of gravitational equations. Conversely, if $f(R)$ gravitational equation hold, the total entropy of the fluid should be extremum. Our work suggests a general and solid connection between $f(R)$ gravity and thermodynamics.}


\begin{document}
\newcommand{\bea}{\begin{eqnarray*}}
\newcommand{\eea}{\end{eqnarray*}}
\newcommand{\bean}{\begin{eqnarray}}
\newcommand{\eean}{\end{eqnarray}}
\newcommand{\eqs}[1]{Eqs. (\ref{#1})}
\newcommand{\eq}[1]{Eq. (\ref{#1})}
\newcommand{\meq}[1]{(\ref{#1})}
\newcommand{\fig}[1]{Fig. \ref{#1}}
\newcommand{\ppa}[2]{\left(\frac{\partial}{\partial #1}\right)^{#2}}
\newcommand{\pp}[2]{\left(\frac{\partial #1}{\partial #2}\right)}
\newcommand{\ppn}[2]{\frac{\partial #1}{\partial #2}}

\newcommand{\tri}{\delta}
\newcommand{\grad}{\nabla}
\newcommand{\pa}{\partial}
\newcommand{\pf}[2]{\frac{\pa #1}{\pa #2}}
\newcommand{\cla}{{\cal A}}

\newcommand{\eqn}{&=&}
\newcommand{\non}{\nonumber \\}
\newcommand{\oh}{\frac{1}{2}}
\newcommand{\bs}{\boldsymbol}
\newcommand{\sgt}{\sqrt{h}}
\newcommand{\hsp}{\hspace{0.1mm}}
\newcommand{\hij}{h_{ij}}
\newcommand{\rt}{R^{(3)}}

\section{Introduction}

Black holes are mysterious and important objects which have been studied for a long time. The mathematical analogy between laws of black hole mechanics and the ordinary laws of thermodynamics leads to the discovery of black hole thermodynamics \cite{b1}. Hawking proved that the black hole behaves like a black body with a temperature which is proportional to its surface gravity, and one quarter of the horizon area is the entropy of the black hole [2, 3]. Since then, black hole thermodynamics has been widely studied, and people believe that it could help us to catch sight of some important fundamental theories.

It is generally believed that the gravity equation is the basic equation of nature. However, Jacobson put forward a new point of view that the Einstein equation is an equation of state \cite{Jacob1995}. Over the years this point has been accepted by more and more people. R. Cai proved that applying the first law of thermodynamics to the apparent horizon of a Friedmann-Robertson-Walker universe, one could derive the Friedmann equations describing the dynamics of the universe \cite{Cai2005}. And Verlinde considered gravity could be explained as an entropic force \cite{Ver2011}. Bravetti presented an example of spacetimes effectively emerging from the thermodynamic limit over an unspecified microscopic theory without any further assumptions \cite{Brav2015}. All these discussion showed that the gravity equation may not be the basic assumption but the thermodynamics relation is.

In fact, before the establish of black hole thermodynamics, Cocke \cite{cocke} proposed maximum entropy principle for self-gravitating fluid. In contrast to black hole, the local thermodynamic quantities of self-gravitating fluid, such as entropy density $s$, energy density $\rho$ and local temperature $T$, are well defined. The presence of gravity only affects the distribution of these local quantities. Roughly speaking, there are two different ways to approach the distribution of the self-gravitating fluid system. One way is called the dynamical method, and the other way is called the thermodynamical method. In dynamical method, using Einstein's equation and applying conservation of matter, people could obtain the distribution of the fluid. But in thermodynamical method, using only constrain Einstein equation and the ordinary thermodynamical relation, then the fluid could be configured such that its total entropy attains a maximum value. For a spherical radiation system, Sorkin, Wald and Zhang \cite{wald81} showed that if the total entropy be an extremum and the Einstein constraint equation holds, then the Tolman-Oppenheimer-Volkoff (TOV) equation of hydrostatic equilibrium can be derived. Gao \cite{gao} generalized SWZ's work to arbitrary perfect fluid in static spherical spacetime and got the TOV equation for this fluid. This issue has been further explored in the past years [11-14]. Recently, series of general proofs of the maximum entropy principle in the case of static spacetime without the spherical symmetry has been completed, included Einstein-Maxwell theory and Lovelock theory [15-18].

However, whether the principle is valid for $f(R)$ gravity is not clear. $f(R)$ gravity is a natural generalization of Einstein gravity. The motivation of modifying generalization of GR comes from high-energy physics for adding higher order invariants to the gravitational action, as well as comes from cosmology and astrophysics for seeking generalizations of GR. As an important modified theories, $f(R)$ theories can explain the accelerated expansion of the universe because it contains the $\alpha R^2$ terms \cite{Starobinsky}. $f(R)$ gravity comes about by a straightforward generalization of the Lagrangian in the Einstein action to become a general function of Ricci scalar $R$, i.e.,
\bean
S=\frac{1}{2\kappa}d^4x\sqrt{-g}f(R)\,,
\eean
where $f$ as a series expansion of $R$
\bean
f(R)=\sum^{n=\infty}_{n=-\infty}\alpha_n R^n\,.
\eean
There are two advantages in $f(R)$ gravity [20, 21]. First, $f(R)$ actions are simple enough to be deal with, but sufficiently general to encapsulate some of the basic characteristics of higher-order gravity. Second, there are some reasons to believe that $f(R)$ gravity is unique among higher-order gravity theories which can avoid Ostrogradski instability. Thus it is interesting to discuss the maximum entropy principle in this generalized gravity theories.

In Sec. II, we discuss the maximum entropy in spherical static spacetime. TOV equation constrains the structure of a spherically symmetric body of isotropic material which is in static gravitational equilibrium. We first briefly review how to obtain the modified TOV equation of $f(R)$ gravity by traditional dynamical method. And then in addition to the constraint equation, we only make use of the ordinary thermodynamical relations to derive the same TOV equation in $f(R)$ gravity. These two different methods obtained the same results shows that the maximum entropy in spherical static spacetime in $f(R)$ theories is valid.

In Sec. III, with the help of Tolman's law, which was also used in [13,15,18], we extend the entropy principle to general static spacetimes without spherical symmetry condition. It should be noticed that, the total entropy $S$ appears to depend on $\delta h_{ab}$ and $\delta R$ (See Appendix A). Fixing the induced metric $h_{ab}$ and scalar curvature $R$ and their first derivatives on the boundary, in order to allow us to use integration by parts and drop the boundary terms. And it also shows that the components of $\delta R$ vanishes automatically. So the vanishing of the components of $\delta h_{ab}$ just gives the spatial components of gravitational equation of $f(R)$ theories. The derivation of this section shows that the maximum entropy is also valid in $f(R)$ theories without spherical symmetry.

Throughout our discussion, symbol $a, b, c$ represent the abstract index notation, $\grad_a$ and $D_a$ are the covariant derivative associated with the metric of spacetime and the covariant derivative associated with the induced metric $h_{ab}$, respectively, and $\Box\equiv\grad^a\grad_a$.

\section{TOV equations obtained by two different methods}
\subsection{Classical dynamical method}
As for the issue of the distribution of fluid, dynamical method is a classical one, and is widely used. First we briefly review how to get TOV equation in $f(R)$ gravity by dynamical method \cite{hR}. In 4-dimensional spacetime the action of the f(R) gravity can be expressed as
\bean
S=\frac{1}{2\kappa}\int d^4x\sqrt{-g}f(R)+S_M\,,
\eean
where $\kappa\equiv 8\pi$, and $S_M$ is the action term of perfect fluid matter. Variation  with respect to the metric gives
\bean
f_R(R)R_{ab}-\frac{1}{2}f(R)g_{ab}-[\grad_{a}\grad_{b}-g_{ab}\Box]f_R(R)=8\pi T_{ab}\,,\label{Tab}
\eean
where
\bean
T_{ab}=-\frac{2}{\sqrt{-g}}\frac{\delta S_M}{\delta g^{ab}}\,,
\eean
and $f_R(R)\equiv\partial f/\partial R$.

For a static perfect fluid system with spherical symmetries in the f(R) gravity, the line element of the spacetime can be written as
\bean
ds^2=-e^{2\phi(r)}dt^2+e^{2\lambda(r)}dr^2+r^2(d\theta^2+\sin^2\theta d\phi^2) \,.
\eean
The energy-momentum tensor of the perfect fluid is
\bean
T_{ab}=\rho u_au_b+p h_{ab}\,,\label{energymom}
\eean
where $\rho$ is the matter density and $p$ is the pressure, and in our chosen coordinate we have $u^a = -e^{\phi(r)}(dt)_a$. The components of the field equations become
\bean
8\pi T_t\hsp^t = 8\pi\rho \eqn \frac{f_R(R)}{r^2}+\frac{e^{-2\lambda}(2r\lambda'-1)} {r^2}f_R(R)+\frac{1}{2}f(R) \non
&& -\frac{R}{2}f_R(R)-e^{-2\lambda}f_R''(R) +e^{-2\lambda}(\lambda'-\frac{2}{r})f_R'(R) \,, \label{rho}
\eean
and
\bean
8\pi T_r\hsp^r = 8\pi p \eqn -\frac{f_R(R)}{r^2}+\frac{e^{-2\lambda}(1+2r\phi')} {r^2}f_R(R)-\frac{1}{2}f(R) \non
&& +\frac{R}{2}f_R(R)+e^{-2\lambda}\frac{2+r\phi'}{r}f_R' \,,\label{p}
\eean
where $' \equiv d/dr$.

To get the TOV equation, we consider Bianchi identity, which leads to energy-momentum conservation equation $\grad_a T^a\hsp_b=0$. The radial component of the conservation equation gives
\bean
p'=-(\rho+p)\phi'\,.
\eean
If we assume
\bean
e^{-2\lambda} \eqn 1-\frac{2m}{r}\,.
\eean
Then the modified TOV equations take the following convenient form
\bean
&& 8\pi p+\frac{2m}{r^3}f_R(R)+\frac{f(R)}{2}-\frac{R}{2}f_R(R)-\frac{2(r-2m)}{r^2}f_R'(R) \non
&& =-\frac{r-2m}{r}\left(\frac{2f_R(R)}{r}+f_R'(R)\right)\frac{p'}{p+\rho}\,. \label{DynaTOV}
\eean
Notice that if the Lagrangian degenerate to Einstein theory, i.e. $f(R)=R$, \eq{DynaTOV} will reduce to the well-known TOV equation in Ref. \cite{TOV}.

\subsection{Thermodynamical method}
In previous subsection, we have shown how to get the TOV equation in $f(R)$ gravity by using dynamical method. Generally, to get the TOV equation, one has to use radial-radial component (\eq{p}) of the gravitational equation. Now, we shall derive TOV equation in $f(R)$ gravity by using thermodynamical method. We only use conservation equation (\eq{rho}) and the extrema of total entropy, admits the condition of the variation of total particle number vanish in static spacetime, to derive TOV equation of $f(R)$ gravity. It implies that the maximum entropy principle contains part of information of gravitational equations in $f(R)$ gravity.

The perfect fluid system satisfies the familiar first law
\bean
dS=\frac{1}{T}dE+\frac{P}{V}dV-\frac{\mu}{T}dN\,, \label{firstlawV}
\eean
where $S$, $E$, $N$ represent the total entropy, energy, and particle number within the volume $V$. Their density variables are denoted by $s$, $\rho$ and $n$, respectively. Applying \eq{firstlawV} to a unit volume, it is easy to find
\bean
ds=\frac{1}{T}d\rho-\frac{\mu}{T}dn\,,
\eean
and the integrated form which be called the Gibbs-Duhem relation
\bean
s=\frac{p+\rho-\mu n}{T}\,.\label{Gibbs}
\eean

Now we apply the maximum entropy principle to this self-gravitating perfect fluid in $f(R)$ gravity. In spherical symmetry case, the total entropy and the total particle number from $r=0$ to $r=r_0$ can be written as \cite{gao}
\bean
S \eqn 4\pi\int_0^{r_0}s(r)\left[1-\frac{2m(r)}{r}\right]^{-1/2}r^2dr\,,\\
N \eqn 4\pi\int_0^{r_0}n(r)\left[1-\frac{2m(r)}{r}\right]^{-1/2}r^2dr\,.
\eean
The variation of total particle number vanishes, which means $\delta N=0$, is a desired condition in maximum entropy principle. Following the standard method of Lagrange multipliers, the equation of variation becomes
\bean
\delta S+\lambda\delta N=0\,.
\eean
Define the "total Lagrangian" by
\bean
L(m,m',n)=s(\rho(m,m'),n)\left(1-\frac{2m(r)}{r}\right)^{-\frac{1}{2}}r^2+\lambda n(r)\left(1-\frac{2m(r)}{r}\right)^{-\frac{1}{2}}r^2\,.
\eean
So the constrained Euler-Lagrange equation is given by
\bean
\frac{\partial L}{\partial n} \eqn 0 \,,\label{plpn}\\
\frac{d}{dr}\frac{\partial L}{\partial m'} \eqn \frac{\partial L}{\partial m}\,. \label{plpm}
\eean
\eq{plpn} yields,
\bean
\frac{\partial s}{\partial n} \eqn \lambda \,.
\eean
By using thermodynamic relation we have
\bean
\mu=\lambda T\,.\label{muT}
\eean

Now consider \eq{plpm}, after some calculation, we have
\bean
\frac{\partial L}{\partial m} \eqn \frac{1}{8\pi T}\left(\frac{2}{r}f_R''+\frac{3}{r^2}f_R'\right)r^2\left(1-\frac{2m}{r}\right)^{-1/2}+(n\lambda+s)r\left(1-\frac{2m}{r}\right)^{-3/2}\,,
\eean
and
\bean
\frac{\partial L}{\partial m'} \eqn \frac{1}{8\pi T}\left(\frac{2}{r^2}f_R(R)+\frac{1}{r}f_R'(R)\right)r^2\left(1-\frac{2m}{r}\right)^{-1/2}\,.
\eean
By substituting these results into \eq{plpm}, together with \eqs{Gibbs} and \meq{muT}, we could obtain the Euler-Lagrangian equation of $m$ in an explicit form
\bean
&& \frac{T'}{8\pi T^2}\left(\frac{2}{r^2}f_R(R)+\frac{1}{r}f_R'(R)\right)r^2 \non
\eqn \frac{1}{8\pi T}\left(-\frac{4}{r^3}f_R(R)-\frac{1}{r^2}f_R'(R)
+\frac{2}{r^2}f_R'(R)+\frac{1}{r}f_R''(R)\right)r^2 \non
&& +\frac{1}{8\pi T}\left(\frac{2}{r^2}f_R(R)+\frac{1}{r}f_R'(R)\right)(2r+rm'-5m)\left(1-\frac{2m}{r}\right)^{-1} \non
&& -\frac{1}{8\pi T}\left(\frac{2}{r}f_R''+\frac{3}{r^2}f_R'\right)r^2-\frac{p+\rho}{T}r\left(1-\frac{2m}{r}\right)^{-1}\,.\label{TprimeT}
\eean
The constraint \eq{muT} yield
\bean
\mu'=\lambda T'\,.
\eean
Considering the fundamental thermodynamic relation
\bean
dp\eqn sdT+nd\mu\,,
\eean
it follows that
\bean
p'=sT'+n\mu'=\frac{T'}{T}(\rho+p) \,.
\eean
Thus we obtain the equation
\bean
&& \frac{p'}{\rho+p}\left(\frac{2}{r^2}f_R(R)+\frac{1}{r}f'(R)\right)r^2(1-\frac{2m}{r}) \non
\eqn \left(-\frac{4}{r^3}f_R(R)+\frac{1}{r^2}f_R'+\frac{1}{r}f_R''\right)r^2\left(1-\frac{2m}{r}\right)+\left(\frac{2}{r^2}f_R+\frac{1}{r}f_R'\right)(2r+rm'-5m) \non
&& -\left(\frac{2}{r}f_R''(R)+\frac{3}{r^2}f_R'(R)\right)r^2(1-\frac{2m}{r})-8\pi pr-8\pi \rho r \,. \label{TherTOV}
\eean
By substituting \eq{rho} into \eq{TherTOV}, finally we get
\bean
&& -\frac{p'}{\rho+p}\left(\frac{2}{r^2}f_R(R)+\frac{1}{r}f'(R)\right)r^2(1-\frac{2m}{r}) \non
\eqn 8\pi p+\frac{f(R)}{2}+\left(\frac{2m}{r^3}-\frac{R}{2}\right)f_R(R)-\frac{2(r-2m)}{r^2}f_R'(R)\,.
\eean
This TOV equation in $f(R)$ gravity is exactly the same as \eq{DynaTOV}.

It is necessary to point out that we only use the maximum entropy principle and the time-time component gravitational equation. No other assumptions are needed. Which tells us that in spherical symmetry case, the extrema of entropy contain part of information of gravitational equation. And this is a direct evidence for the fundamental relationship between gravitation and thermodynamics.

\section{Maximum entropy principle for general static spacetime in $f(R)$ gravity}

In previous sections, we got TOV equation by using dynamical and thermodynamic methods, respectively. But all these discussions based on static spherical symmetry condition. What we want to know is that whether the maximum entropy principle is valid for general static spacetime in $f(R)$ gravity.

Consider a general static spacetime ($M$, $g_{ab}$) and $\Sigma$ as a three dimensional hypersurface denoting a moment of the static observers. Let $C$ be a region on $\Sigma$ with a boundary $\bar C$ and $h_{ab}$ be the induced metric on $\Sigma$, and $R$ be the scalar curvature of spacetime, respectively. We assume that the Tolman's law holds in $C$, like\cite{fang1}, which means that the local temperature $T$ of the fluid satisfies
\bean
T\chi=T_0\,, \label{Tchi}
\eean
where $\chi$ is the red-shift factor for static observers and $T_0$ is a constant which can be viewed as the red-shift temperature of the fluid. Without loss of generality, we take $T_0=1$ in Tolman's law.  Assume that the constraint equation is satisfied in $C$. We also need $h_{ab}$, $R$ and their first derivatives are fixed on $\bar C$, which allow us to use integration by parts and drop the boundary term. Then, it can be proved that the other components of gravitational equation are implied by the extrema of the total fluid entropy for all variations of data in $C$.

We consider a general perfect fluid as discussed in previous section. The induced metric on $\Sigma$ is given by
\bean
h_{ab}=g_{ab}+u_au_b\,.
\eean
The stress-energy tensor $T_{ab}$ takes the form of \eq{energymom}. And from conservation law $\grad_aT^{ab}=0$, we can show that
\bean
\grad_ap=-(\rho+p)A_a\,,
\eean
where $A_a$ is the four-acceleration of the observer. Since
\bean
u^a\eqn \frac{\xi^a}{\chi}\,,
\eean
where $\xi^a$ is the Killing vector. One can show that
\bean
A_a=\grad_a\chi/\chi\,, \label{Achi}
\eean
and thus
\bean
\grad_a p=-(\rho+p)\grad_a\chi/\chi\,.\label{Tchi1}
\eean
Meanwhile, the local first law can also be expressed in the form
\bean
dp=sdT+\mu dn\,. \label{Tchi2}
\eean
Comparing with \eqs{Tchi1} and \meq{Tchi2}, and applying the Tolman's law \eq{Tchi} and the Gibbs-Duhem relation \eq{Gibbs}, we finally have
\bean
\mu\chi=\mu/T=constant\,.
\eean
If we choose energy density $\rho$ and particle number density $n$ as two independent thermodynamical variables. Then in static spacetime, the total entropy $S$ is an integral of the entropy density $s$ over the region $C$ on $\Sigma$,
\bean
S=\int_C\sgt s(\rho,n)\,.
\eean
where $h$ is the determinant of $h_{ab}$. Thus, the variation of the total entropy can be written as
\bean
\delta S=\int_Cs\delta\sgt+\sgt\delta s\,.
\eean
Using the local first law of thermodynamics,
\bean
Tds=d\rho-\mu dn\,,
\eean
we find
\bean
\delta S \eqn \int_C s\delta\sgt+\sgt\left(\frac{\partial s}{\partial\rho}\delta\rho+\frac{\partial s}{\partial n}\delta n\right) \non
\eqn \int_Cs\delta\sgt+\sgt\left(\frac{1}{T}\delta\rho-\frac{\mu}{T}\delta n\right)\,.
\eean
Notice that $\mu/T$ is constant, which can be taken out of the integral. It is natural to require the total number of particles
\bean
N=\int_C\sgt n\,,
\eean
to be invariant, which leads to
\bean
\int_C\sgt\delta n=-\int_C n\delta\sgt\,.
\eean
So we have
\bean
\delta S \eqn \int_C\left(s+\frac{n\mu}{T}\right)\delta\sgt+\sgt\frac{1}{T}\delta\rho \non
\eqn \int_C\frac{\rho+p}{T}\delta\sgt+\sgt\frac{1}{T}\delta\rho \,.
\eean
Using \cite{waldbook}
\bean
\delta\sgt=\frac{1}{2}\sgt h^{ab}\delta h_{ab}\,,
\eean
and denoting
\bean
\delta S=\int_C\delta L\,,
\eean
we have
\bean
\delta L=\frac{1}{2}\frac{\rho+p}{T}\sgt h^{ab}\delta h_{ab}+\sgt\frac{1}{T}\delta\rho\,.\label{detL}
\eean
After some calculation, we get the exactly express of the term $\sgt\delta\rho/T$ in Appendix A. By substituting \eq{detLrho} into \eq{detL}, we have
\bean
\delta L \eqn \frac{\rho+p}{2T}\sgt h^{ab}\delta h_{ab}+\delta L_{\rho} \non
\eqn  \frac{\rho+p}{2T}\sgt h^{ab}\delta h_{ab}+\frac{\sgt}{16\pi} D^bD^a(\chi f_R(R))\delta h_{ab}
-\frac{\sgt}{16\pi}D_c[h^{ab}D^c(\chi f_R(R))]\delta h_{ab} \non
&& -\frac{\sgt}{16\pi}f_R(R)\chi\rt\hsp^{ab}\delta h_{ab}-\frac{\sgt}{8\pi}D^a\chi(D^bf_R(R))\delta h_{ab}+\frac{\sgt}{16\pi} h^{ab}D_c(\chi D^cf_R(R))\delta h_{ab}\,.\nonumber\\
\label{detL2}
\eean

So far, we only applied the constraint equation $G_{00}=T_{00}$. This shows explicitly that $\delta S$ is determined only by the variation of $h_{ab}$. Since $\delta S=0$ in our assumption, it is clear shows that $\delta L=0$. Together with \eq{rhogeneral} and Tolman's law \eq{Tchi}, it is easily find
\bean
\frac{8\pi p h^{ab}}{T}\delta h_{ab} \eqn \frac{h^{ab}}{T}\delta h_{ab}\left[h^{ac}h^{bd}R_{cd}f_R(R)-\frac{1}{2}f(R)h^{ab}-h^{ac}h^{bd}\grad_c\grad_d f_R(R)+h^{ab}\Box f_R(R)\right] \non
&& -\frac{1}{2}(P_1^{ab}+P_2^{ab})\delta h_{ab}\,, \label{proj}
\eean
where $P_1$ denotes the terms parallel to $h^{ab}$,
\bean
P_1^{ab} \eqn \chi h^{ab}\left[f_R(R)R_{cd}u^cu^d+\frac{1}{2}f(R)-u^cu^d\grad_c\grad_d f_R(R)-\Box f_R(R)\right] \non
&-&\chi\frac{1}{2}f(R)h^{ab}+h^{ab}\chi\Box f_R(R)-h^{ab}D_cD^c(\chi f_R(R))+h^{ab}D_c(\chi D^cf_R(R))\,, \non
\eean
and $P_2$ denotes the other terms,
\bean
P_2^{ab} \eqn \chi[f_R(R)R_{cd}h^{ac}h^{bd}-h^{ac}h^{bd}\grad_c\grad_d f_R(R)]+D^bD^a(\chi f'(R)) \non
&& -f'(R)\chi\rt\hsp^{ab}-2D^a\chi(D^bf_R(R))\,.
\eean
We show that $P_1^{ab}$ and $P_2^{ab}$ are vanish in Appendix B. So \eq{proj} gives the projection of gravitational equation
\bean
8\pi p \eqn h^{ac}h^{bd}R_{cd}f_R(R)-\frac{1}{2}f(R)h^{ab}-h^{ac}h^{bd}\grad_c\grad_d f_R(R)+h^{ab}\Box f_R(R)\,.
\eean

In the above proof, under the fixed boundary condition, we only used the constraint \eq{rhogeneral} to derive \eq{detL2}. By applying the extrema of entropy $\delta S=0$, we obtained the spatial components of gravitational equation. It is easy to check that the proof is reversible. From the spatial components of gravitational equation, one can show $\delta L=0$ in \eq{detL2}.

\section{Summary and discussions}
The TOV equation is an important equation for self-gravitating system. In this paper, we first briefly review how to get modified TOV equation in $f(R)$ gravity by dynamical method. By applying both time-time and radial-radial gravitational equations, and the radial component constraint equation, one can show that the general TOV equation in $f(R)$ gravity is given by \eq{DynaTOV}. Then, by applying the maximum entropy principle to a general self-gravitating fluid, we have obtained the same TOV equation of hydrostatic equilibrium by thermodynamical method. We should point out that, to derive the TOV equation by the extrema of the total entropy $S$, we only need the constraint equation and fix the total particle number of the system. Comparing with these two different processes, it tells us that the maximum entropy principle is valid in spherical symmetry case in $f(R)$ theories.

We also prove that the maximum entropy principle of perfect fluid is valid to case without any symmetry on spacelike hypersurface. Arbitrary select a region, consider the fluid obeys Tolman's law and the constraint equation is satisfied in this region. Then fix the induced metric $h_{ab}$ and scalar tensor $R$ and their first derivative on the boundary of selected region. Our calculation shows that $\delta S$ appears to depend on $\delta h_{ab}$ and $\delta R$, and the component of $\delta R$ always vanish. It shows that the variation of total entropy in this region is exactly linear to $\delta h_{ab}$. The assumption of $\delta S=0$ means the coefficient of $\delta h_{ab}$ vanish, which imply that the other components of gravitational equation satisfied.

Till now, the maximum entropy principle has been set up in series of theories. Although it is unclear whether this principle could be used for a general Lagrangian of gravity theories which needs further study, we consider that there may exist some deeper connection between gravity theories and thermodynamics.

\section*{Acknowledgements}
We thank S. Gao and X. He for many helpful discussions. This work is supported by the National Natural Science Foundation of China under Grant No. 11475061;  the SRFDP under Grant No. 20114306110003; the Open Project Program of State Key Laboratory of Theoretical Physics, Institute of Theoretical Physics, Chinese Academy of Sciences, China Grant No. Y5KF161CJ1. Xiongjun Fang is also supported by China Postdoctoral Science Foundation funded project.

\section{Appendix}
\subsection{Appendix A: Detail calculation of $\delta\rho$}
In this Appendix, we will show the detailed calculation of $\delta \rho$ in \eq{detL}. First, note that the extrinsic curvature of $\Sigma$, is defined by
\bean
\hat B_{ab}=h_a\hsp^ch_b\hsp^d\grad_du_c\,.
\eean
It is straightforward to show
\bean
\hat B_{ab}=\grad_b u_a+A_au_b\,,
\eean
where $A_a$ is the four-acceleration of the observer . And in static spacetime we have
\bean
\hat B_{ab}=0\,,
\eean
so we get a very helpful formula in our calculation,
\bean
\grad_b u_a=-A_a u_b\,.\label{uaab}
\eean

Then, we show \cite{waldbook} that the curvature $\rt_{abc}\hsp^d$ of hypersurface $\Sigma$ is related to the spacetime curvature $R_{abc}\hsp^d$ by
\bean
\rt_{abc}\hsp^d=h_a\hsp^f h_b\hsp^g h_c\hsp^k h_j\hsp^d R_{fgk}\hsp^j\,,
\eean
From this definition, it is easy to find
\bean
\rt_{ab}=R_{ab}+R_{aeb}\hsp^lu^eu_l+R_{fb}u^fu_a+R_{ak}u^ku_b+u_au_bR_{fk}u^fu^k\,,\label{r3ab}
\eean
and
\bean
\rt=R+2R_{ab}u^au^b\,.\label{r3}
\eean
Meanwhile, from \eq{r3ab} we also have
\bean
\rt\hsp^{ab}=\rt_{cd}h^{ac}h^{bd}=R^{ab}+R_{aeb}\hsp^lu^eu_lh^{ac}h^{bd}\,.\label{r3abup}
\eean
From \eq{energymom}, one can easily get the Hamiltonian constraint of the theory,
\bean
8\pi\rho=u^au^bR_{ab}f_R(R)+\frac{1}{2}f(R)-u^au^b\grad_a\grad_b f_R(R)-\Box f_R(R)\,. \label{rhogeneral}
\eean
Using \eq{r3}, the variation of $\rho$ is
\bean
8\pi\delta\rho \eqn \delta\left[u^au^bR_{ab}f_R(R)+\frac{1}{2}f(R)-u^au^b\grad_a\grad_b f_R(R)-\Box f_R(R)\right] \non
\eqn \frac{1}{2}\delta[(\rt-R)f_R(R)]+\frac{1}{2}\delta f(R)-\delta[u^cu^d\grad_c\grad_d f_R(R)]-\delta[\Box f_R(R)] \non
\eqn \frac{1}{2}f_R(R)\delta\rt+\frac{1}{2}(\rt-R)f_{RR}(R)\delta R \non
&& -\delta[u^cu^d\grad_c\grad_d f_R(R)]-\delta[\Box f_R(R)]\,,
\eean
where we denote $f_{RR}(R)\equiv\frac{\partial^2f}{\partial R^2}$, and we use $\delta f(R)=f_R(R)\delta R$. So $\delta\rho$ term in \eq{detL} becomes
\bean
\delta L_{\rho} \eqn \sgt\frac{\delta\rho}{T}  \non
\eqn \frac{\sgt}{16\pi T}f_R(R)\delta\rt+\frac{\sgt}{16\pi T}(\rt-R)f_{RR}(R)\delta R  \non
&& -\frac{\sgt}{8\pi T}\delta[u^au^b\grad_a\grad_b f_R(R)]-\frac{\sgt}{8\pi T}\delta[\Box f_R(R)]\,.  \label{detrho}
\eean

Now, we deal with terms of $\delta L_{\rho}$, respectively. Notice that the Tolman's law \eq{Tchi} is used in our calculation. The first term of \eq{detrho} can be calculate as
\bean
8\pi\delta L_{\rho1} \eqn  \frac{\sgt}{2T}f_R(R)\delta\rt = \frac{1}{2}\chi\sgt f_R(R)(h^{ab}\delta\rt_{ab}+\rt_{ab}\delta h^{ab})\,.
\eean
The standard calculation of $\delta\rt_{ab}$ yields
\bean
h^{ab}\delta\rt_{ab}=\sgt D^a(D^b\delta h_{ab}-h^{bc}D_a\delta h_{bc})\,,
\eean
Using integration by parts and dropping boundary terms, we have
\bean
8\pi\delta L_{\rho1} \eqn \frac{1}{2}\sgt D^bD^a(\chi f_R(R))\delta h_{ab}-\frac{1}{2}\sgt D_c[h^{ab}D^c(\chi f_R(R))]\delta h_{ab} \non
&& -\frac{1}{2}f_R(R)\chi\sgt\rt\hsp^{ab}\delta h_{ab}\,. \label{detrho1}
\eean

The third term of \eq{detrho} can be calculate as
\bean
8\pi\delta L_{\rho3} \eqn  -\frac{\sgt}{T}\delta[u^au^b\grad_a\grad_b f_R(R)]\,,
\eean
and the fourth term of \eq{detrho} can be written as
\bean
8\pi\delta L_{\rho4}\eqn-\frac{\sgt}{T}\delta[\Box f'(R)] = -\frac{\sgt}{T}\delta(\grad_a\grad^af'(R))\,.
\eean

Note that
\bean
D_aD^af_R(R) \eqn h^{ab}\grad_a\grad_b f_R(R) \non
\eqn \grad_a\grad^a f_R(R)+u^au^b\nabla_a\nabla_b f_R(R)\,,
\eean
and
\bean
\delta(D_aD^af_R(R))=D_a\delta(D^af_R(R))+\delta C^{c}\hsp_{ac}D^af_R(R)\,,
\eean
where
\bean
\delta C^{c}\hsp_{ac}=\frac{1}{2}h^{cb}D_a\delta h_{cb}\,.
\eean

So we have
\bean
&& 8\pi\delta L_{\rho3}+8\pi\delta L_{\rho4}\non
\eqn-\frac{\sgt}{T}[D_a\delta(D^af_R(R))+\delta C^{c}\hsp_{ac}D^af_R(R)]\non
\eqn -\sgt D^a\chi(D^bf_R(R))\delta h_{ab}-\sgt (D^aD_a\chi)f_{RR}(R)\delta R \non
&& +\frac{1}{2}\sgt h^{ab}D_c(\chi D^cf_R(R))\delta h_{ab}  \,. \label{detLrho4}
\eean
Considering the second term of \eq{detrho}, together with \eqs{detLrho4} and \meq{r3}, we denote all the terms contained $f_{RR}(R)\delta R$ by $\delta L_{\rho\delta R}$, then
\bean
8\pi\delta L_{\rho\delta R} \eqn -\sgt(D^aD_a\chi)f_{RR}(R)\delta R+\frac{\sgt}{2}\chi(\rt-R)f_{RR}\delta R \non
\eqn -\sgt(D^aD_a\chi)f_{RR}(R)\delta R+\frac{\sgt}{2}\chi u^au^bR_{ab} f_{RR}\delta R\,.
\eean

We can calculate that
\bean
\chi R_{ab}u^au^b \eqn R_{acb}\hsp^cu^au^b \non
\eqn -\chi u^a\grad_a\grad_c u^c+\chi u^a\grad_c\grad_a u^c \non
\eqn \chi u^a\grad_c(-u_aA^c) = \chi\grad_aA^a\,. \label{Rabuaub}
\eean
And we know
\bean
\grad_aA^a \eqn A_aA^a+D_aA^a\,.\label{gradA}
\eean
So the $f_{RR}(R)\delta R$ terms vanish,
\bean
\delta L_{\rho\delta R}=0\,.
\eean

At last, the calculation of $\delta\rho$ term of \eq{detL} shows
\bean
\delta L_{\rho} \eqn \frac{\sgt}{16\pi} D^bD^a(\chi f_R(R))\delta h_{ab}-\frac{\sgt}{16\pi} D_c[h^{ab}D^c(\chi f_R(R))]\delta h_{ab}-\frac{\sgt}{16\pi}f_R(R)\chi\rt\hsp^{ab}\delta h_{ab}  \non
&& -\frac{\sgt}{8\pi}\chi A^aD^bf_R(R)\delta h_{ab}-\frac{\sgt}{8\pi} D^a\chi(D^bf_R(R))\delta h_{ab}+\frac{\sgt}{16\pi} h^{ab}D_c(\chi D^cf_R(R))\delta h_{ab} \non
&& +\frac{\sgt}{8\pi}\chi A^a(D^b f_R(R))\delta h_{ab} \non
\eqn \frac{\sgt}{16\pi} D^bD^a(\chi f_R(R))\delta h_{ab}-\frac{\sgt}{16\pi}D_c[h^{ab}D^c(\chi f_R(R))]\delta h_{ab}-\frac{\sgt}{16\pi}f_R(R)\chi\rt\hsp^{ab}\delta h_{ab}  \non
&& -\frac{\sgt}{8\pi}D^a\chi(D^bf_R(R))\delta h_{ab}+\frac{\sgt}{16\pi} h^{ab}D_c(\chi D^cf_R(R))\delta h_{ab}\,. \label{detLrho}
\eean
This result tells us that the variation of $\rho$ term is exactly determined only by the variation of induced metric $\delta h_{ab}$.

\subsection{Appendix B: Calculation of $P_1^{ab}$ and $P_2^{ab}$}
In this Appendix, we give the detailed calculation of $P_1^{ab}$ and $P_2^{ab}$. We first calculate
\bean
P_1^{ab} \eqn \chi h^{ab}\left[f_R(R)R_{cd}u^cu^d+\frac{1}{2}f(R)-u^cu^d\grad_c\grad_d f_R(R)-\Box f_R(R)\right] \non
&& -\frac{1}{2}\chi f(R)h^{ab}+h^{ab}\chi\Box f_R(R)-h^{ab}D_cD^c(\chi f_R(R)) \non
&& +h^{ab}D_c(\chi D^cf_R(R)) \non
\eqn h^{ab}[\chi f_R(R)R_{cd}u^cu^d-\chi u^cu^d\grad_c\grad_d f_R(R) \non
&& -D_cD^c(\chi f_R(R))+D_c(\chi D^cf_R(R))]\,. \label{p1}
\eean
In static spacetime, we have
\bean
&& -\chi u^cu^d\grad_c\grad_d f_R(R) \non
\eqn -\chi u^c\grad_c[u^d\grad_d f_R(R)]+\chi u^c(\grad_cu^d)\grad_d f_R(R) \non
\eqn \chi A^d D_d f_R(R)\,. \label{p11}
\eean
And notice \eqs{Rabuaub} and \meq{gradA},
\bean
\chi R_{cd}u^cu^d \eqn \chi\grad_c A^c \non
\eqn \chi(A_cA^c+D_cA^c)\,.\label{p12}
\eean
By substituting \eqs{p11} and \meq{p12} in \eq{p1},
\bean
P_1^{ab} \eqn h^{ab}[\chi(A_cA^c+D_cA^c)f_R(R)+\chi A^d D_d f_R(R) \non
&& -D_cD^c(\chi f_R(R))+D_c(\chi D^cf_R(R))] \non
\eqn h^{ab}[A^c(D_c\chi)f_R(R)+\chi (D_cA^c)f_R(R)-(D_cD^c\chi)f_R(R)] \non
\eqn 0\,.
\eean

Then, we calculate $P_2^{ab}$,
\bean
P_2^{ab} \eqn \chi[f_R(R)R_{cd}h^{ac}h^{bd}-h^{ac}h^{bd}\grad_c\grad_d f_R(R)]+D^bD^a(\chi f_R(R)) \non
&& -f_R(R)\chi\rt\hsp^{ab}-2D^a\chi(D^bf_R(R))\,.
\eean
Applying \eq{r3abup}, we have
\bean
P_2^{ab} \eqn \chi f_R(R)R_{cd}h^{ac}h^{bd}-\chi D^aD^bf_R(R)+D^aD^b(\chi f_R(R)) \non
&& -\chi f_R(R)h^{ac}h^{bd}R_{cd}-\chi f_R(R)R_{ced}\hsp^lu^eu^lh^{ac}h^{bd} \non
&& -2D^a\chi(D^bf_R(R))  \non
\eqn f_R(R)(D^aD^b\chi-\chi R_{ced}\hsp^lu^eu^lh^{ac}h^{bd}) \,.
\eean
Using \eq{uaab}, one can show that
\bean
R_{ced}\hsp^lu^eu_lh^{ac}h^{bd} \eqn u^eh^{ac}h^{bd}(\grad_c\grad_eu_d-\grad_e\grad_cu_d) \non
\eqn u^eh^{ac}h^{bd}[\grad_c(-u_e A_d)-\grad_e(-u_c A_d)] \non
\eqn h^{ac}h^{bd}(\grad_c A_d+A_c A_d)=D^aA^b+A^aA^b\,,
\eean
and
\bean
\chi D^aA^b+\chi A^aA^b=D^aD^b\chi\,.
\eean
Hence,
\bean
P_2^{ab}=0
\eean

\end{document}